\documentstyle[twoside,fleqn,espcrc2,epsf]{article}

\def\dt{\delta\tau}                 
\def\pacc{P_{\hbox{\rm\tiny acc}}}  
\def\dH{{\delta H}}                 
\def\Dd{\Delta\delta}               
\def\dH{{\delta H}}                 

\newcommand{\AmS}{{\protect\the\textfont2
  A\kern-.1667em\lower.5ex\hbox{M}\kern-.125emS}}

\title{Non-Reversibility of Molecular Dynamics Trajectories}

\author{R.~G.~Edwards, Ivan Horv\'ath \thanks{{The speaker. Work
supported by the DOE under Grant Nos.~DE-FG05-85ER250000 and
DEFG05-92ER40742.}}, and A.~D.~Kennedy
\address{SCRI, Florida State University, Tallahassee, FL 32306-4052, USA}}
       
\begin{document}

\begin{abstract}
We study the non-reversibility of molecular dynamics trajectories
arising from the amplification of rounding errors. We analyse the causes
of such behaviour and give arguments, indicating that this
does not pose a significant problem for Hybrid Monte Carlo 
computations. We present data for pure $SU(3)$ gauge theory and for
QCD with dynamical fermions on small lattices to illustrate and
to support some of our ideas.

\end{abstract}


\maketitle

The theory of the Hybrid Monte Carlo (HMC) algorithm \cite{duane87}
assumes the exact reversibility of its molecular dynamics (MD)
trajectories. Leapfrog integration guarantees this unless the initial
conjugate gradient (CG) vector is chosen in time asymmetric way
or finite precision arithmetic is used. While the first
condition is easily ensured in practice by using a fixed starting
vector for every CG inversion, all numerical computations carried out
using floating point arithmetic are subject to rounding errors.

These rounding errors are normally not considered dangerous unless
they are exponentially amplified. Indeed, without such an
amplification, the time cost of reducing the error to some preset value
grows only logarithmically with the number of arithmetic operations
involved in the computation. This is a very small correction to the
growth of the cost of the HMC algorithm as the volume and correlation
length of the system are increased.
 
Exponential amplification will occur whenever nearby MD trajectories
diverge from one another exponentially, i.e., when the MD evolution
becomes unstable. There are two distinct mechanisms leading to such
an instability \cite{edwards96}. First, this is typical for nonlinear
equations in the chaotic regime. In fact, the existence of a positive leading
Liapunov exponent for the MD equations of pure $SU(2)$ lattice gauge
theory was proposed in Ref.~\cite{jansen95a}. The second possibility is that
the result of the discrete integration scheme diverges exponentially from
the true solution. This instability should grow with the number of
integration steps and is thus expected to have characteristic time
scale shorter than the one associated with intrinsic chaos. Our  
numerical results confirm this.

The integration instability can be analysed in the context of free
field theory \cite{edwards96}. In fact, the behaviour of a single mode
with frequency $\omega$ already reveals all the essential features.
One can show that the instability accompanied by the exponentially
decaying acceptance rate ($\pacc\sim e^{-\nu\tau}$) occur when
$\omega\dt\ge 2$. Here $\tau$ and $\dt$ are the trajectory length and the
integration step size respectively. In Fig.~1 the '$\sigma=0$' line
shows the characteristic exponent $\nu$ as a function of $\dt$
with $\omega$ fixed to unity. Note the sharp ``wall'' arising at $\dt=2$,
where the instability sets in.

Qualitatively similar behaviour is observed for the case of many stable
modes \cite{edwards96} . The onset of instability is determined
by the highest frequency mode and occurs when $\omega_{max}\dt=2$.
In order to keep the acceptance rate constant for free field theory
as the lattice volume $V\to\infty$, we must decrease $\dt$ so
that $V\dt^4$ stays fixed. Consequently, the instabilities go away
as we approach the thermodynamic limit. In this sense the
leapfrog instability is a {\it finite volume effect}.

In interacting field theory the notion of independent modes loses its
meaning. On the other hand, accepting the standard assumption that 
it can be still useful to think in these terms for asymptotically
free field theories at short distances, it is quite plausible to expect
similar scenario there too. The forces acting on the highest frequency
mode due to the other modes will fluctuate in some complicated
way however, and so
\newpage

\begin{figure}[h]
\vskip -0.1in
\epsfxsize=1.0\hsize
\epsfysize=0.37\vsize
\epsffile{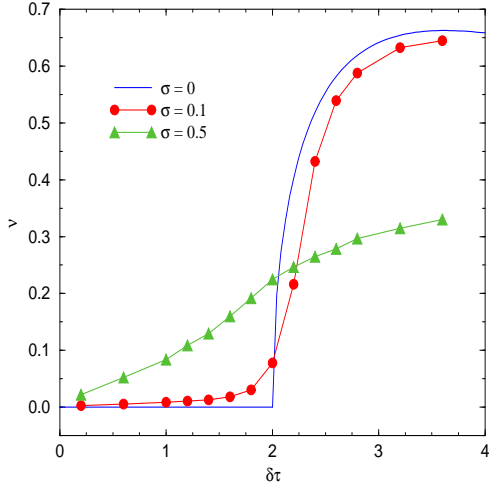}
\vskip -0.56in
\caption{Characteristic exponent for the fluctuating-frequency harmonic 
oscillator model. The frequency fluctuates around $\omega=1$.}
\vskip -0.26in
\end{figure}
\noindent
we expect that the ``wall'' at $\omega_{\max}\dt=2$ will get smeared out.
This is illustrated for the simple model of a harmonic oscillator
whose frequency is randomly chosen from a Gaussian distribution
with mean~$\omega$ and standard deviation~$\sigma$ before each MD step.
The numerical results shown in Fig.~1 confirm that the ``wall'' in
this model does indeed spread out.

Equipped with the above qualitative picture, we have studied 
reversibility numerically for $SU(3)$ gauge theory both in the pure gauge 
and dynamical fermion cases \cite{edwards96} (see also related
work \cite{liu96}). We evolved a typical equilibrium configuration
${U}$ using leapfrog equations for some time $\tau$, then reversed
the momenta and evolved it again for the same amount of time to get
the configuration ${U^\prime}$. Deviations from
reversibility were measured by
    \begin{equation}
       \|\Dd U\|^2 \equiv
       \sum_{x,\mu}\sum_{a,b}|U_{x,\mu}^{a,b}-U_{x,\mu}^{\prime a,b}|^2,
    \end{equation}
but we also recorded the change of energy at the end of the trajectory
($\dH$) and at the end of the reversed trajectory ($\Dd H$).

In Fig.~2 we collected a typical set of data from one pure gauge
configuration. The top and bottom graphs clearly show the integration
instability ``wall'' at $\dt\approx0.6$, which has spread out as
expected. At the same time the middle graph indicates that as we reach
the ``wall'' $\dH=O(10^3)$, so the integration instabilities are
of no practical importance for this system. Note however
the case of the unreasonably long trajectory ($\tau=40$), 
where the reversibility is lost while $\dH$ is very small implying a
good acceptance rate.

\begin{figure}[h]
\vskip -0.23in
\epsfxsize=1.0\hsize
\epsfysize=0.37\vsize
\epsffile{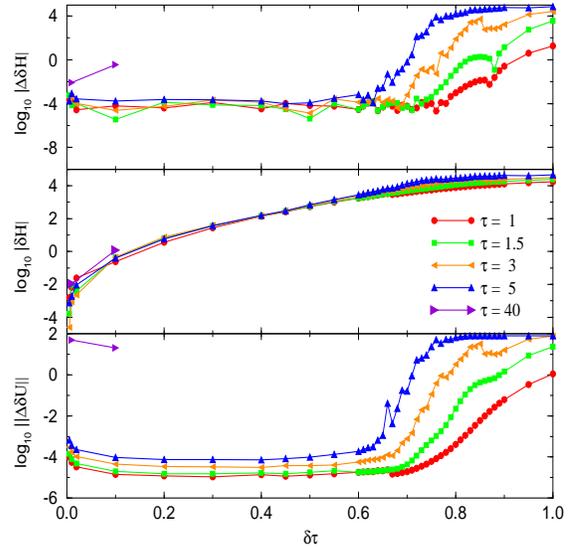}
\vskip -0.40in
\caption[dt-ddu-dh-ddh]{Results for pure $SU(3)$ gauge theory with $\beta=5.7$
    on a $4^4$ lattice as a function of $\dt$.}
\vskip -0.23in
\end{figure}

When ploted as a function of $\tau$, all of our data show a clear
exponential instability in $\|\Dd U\|$. We extracted a characteristic
exponent~$\nu$ ($\|\Dd U\|\sim e^{\nu\tau}$) and show the results
in Fig.~3. Note the same qualitative behaviour we observed for
the toy model in Fig.~1 except that the integration instability ``wall''
appears at different values of $\dt$. This probably just reflects
the different highest frequencies of these systems. In case of full QCD,
the pseudofermions produce a force of the order of the inverse
lightest fermionic mass thus giving the highest relevant frequency
when simulating close to $\kappa_c$. This is reflected in the bottom
graph where the integration instability appears at very small $\dt$.

Notice also that the characteristic exponent does not approach zero for
small $\dt$, which confirms the existence of chaotic continuous
time dynamics. Unlike the integration instability, the intrinsic chaos
can not be controlled by adjusting $\dt$. Moreover, accepting the
standard hypothesis that the trajectory length should be scaled
proportionally to the correlation length in order to reduce the
critical slowing down, non-reversibility might cause problems when
simulating closer to the continuum limit.

\begin{figure}[h]
\vskip 0.01in
\epsfxsize=1.0\hsize
\epsfysize=0.36\vsize
\epsffile{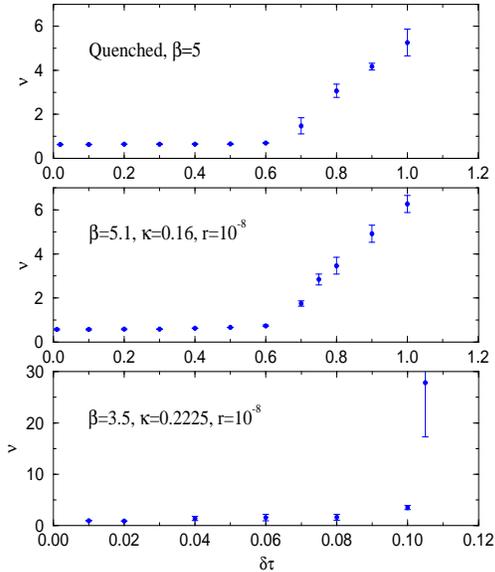}
\vskip -0.43in
\caption{Characteristic exponent for pure $SU(3)$ gauge theory (top),
QCD with heavy dynamical Wilson quarks (middle), and QCD with
light dynamical Wilson quarks (bottom). The data was measured on three
independent configurations.}
\vskip -0.27in
\end{figure}

However, our numerical analysis indicates a strong $\beta-$dependence
of the exponent $\nu$, characterizing the intrinsic chaos. Indeed,
Fig.~4 shows this for $SU(3)$ pure gauge theory on $4^4$ and
$8^4$ lattices. These results can be qualitatively understood if we
hypothesize that chaos is not only a property of this continuous
time evolution, but is also a property of the underlying continuum
field theory. This would suggest that $\nu$ scales like a physical
quantity. At small $\beta$ the lattice theory is in the strong
coupling regime and does not obey the asymptotic scaling behaviour.
At large $\beta$ the system is in a tiny box and is thus in the
deconfined phase. The finite temperature phase transition at
$N_T=4$ occurs near $\beta=5.7$, suggesting that the scaling region is in
the vicinity of this value for our lattices. We have fitted our
$8^4$ data at $\beta=5.4,5.5,5.6,5.7$ to the one loop asymptotic
scaling form $\nu = c e^{-\beta/12\beta_0}$, with
$\beta_0=(11-{2\over3}n_f)/16\pi^2$ and with constant $c$ being the
only free parameter.

\begin{figure}[h]
\vskip -0.41in
\epsfxsize=1.0\hsize
\epsfysize=0.38\vsize
\epsffile{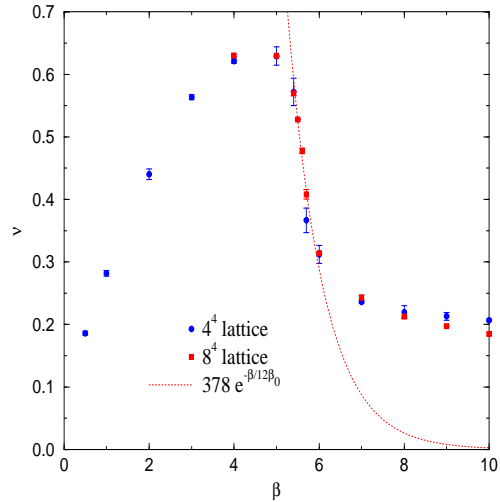}
\vskip -0.58in
\caption{Characteristic exponent as a function of $\beta$ for
pure $SU(3)$ gauge theory. The data was measured on three $4^4$
and two $8^4$ configurations.}
\vskip -0.26in
\end{figure}

The resulting fit, shown in Fig.~4 is surprisingly good,
suggesting that our hypothesis might indeed be correct.
This would mean that the characteristic exponent
is constant when measured in ``physical'' units, that is $\nu\xi$
would be constant as $\xi\to\infty$. If this is the case, then tuning
the HMC algorithm by varying the trajectory length proportionally to
the correlation length does not lead to any change in the
amplification of rounding errors as we simulate closer to the
continuum limit.

\end{document}